\documentclass[aps,preprint,pre,runinaddress,showpacs]{revtex4}
\usepackage{eurosym}
\usepackage{amssymb}
\usepackage{amsmath}
\usepackage{graphicx}
\usepackage{bm}

\begin{document}

\title{Structure and Correlations for Harmonically Confined Charges}
\author{Jeffrey Wrighton and James Dufty }
\affiliation{Department of Physics, University of Florida, Gainesville, FL 32611}
\date{\today}

\begin{abstract}
Coulomb charges confined by a harmonic potential display a rich structure at
strong coupling, both classical and quantum. A simple density functional
theory is reviewed showing the essential role of correlations in forming
shell structure and order within the shells. An overview of previous
comparisons with molecular dynamics and Monte Carlo simulations is summarized and extended.
\end{abstract}

\maketitle

\section{Introduction}

\label{sec1}The problem considered here is the structure and correlation
among $N$ equal Coulomb charges confined by an external harmonic potential.
It is a generalization of the Thomson problem \cite{1} for charges confined
to the surface of a sphere, posed 117 years ago, to three dimensions and
finite temperatures. The ground state for the harmonic confinement is
well-studied, exposing a rich shell structure (distribution of particles
localized about well-defined radii with localization on each radius similar
to those of the single sphere Thomson problem). Within classical mechanics,
these ground state results have been quantified in detail via shell models,
molecular dynamics simulation (MD), and Monte Carlo simulation (MC) \cite{2,3,4,5,6,7}. They are also realized experimentally for dusty plasmas. The
corresponding results for confined charges at finite temperatures is the
extension described here. We provide here a summary of our theoretical work
in collaboration with the Bonitz group at the Institut fur Theoretische
Physik und Astrophysik, Christian-Albrechts Universitat, Kiel \cite{8,9,10,11}.
In addition we describe the method for extension to quantum theory and
list some remaining outstanding problems.

The primary effect of temperature on the classical ground state shell models
is to broaden the sharp shell structure and smooth their angular
distribution due to thermal motion. The governing parameters are the Coulomb
coupling constant $\Gamma $ (ratio of Coulomb to thermal energies of a pair)
and the average number of charges $\overline{N}$. The number of shells is
fixed by $\overline{N}$ while their relative resolution (sharpness) is
determined by $\Gamma $. The zero temperature ground state results from
shell models are recovered in the limit of large $\Gamma .$ A simple
approximate density functional theory described below is able to capture
these results quantitatively, in comparison with those from Monte Carlo
simulations. It is based on approximating correlations among the charges in
the trap by those in the uniform one component plasma. The close
relationship of correlations in these two quite different systems has been
confirmed by MD simulation \cite{10,11}.

At still larger $\Gamma $, corresponding to lower temperatures, the
rotational invariance of the fluid phase is broken and the particles within
each shell become localized about sites close to those of the Thomson
problem for a single shell. Those localized domains are approximated here by
Gaussian distributions centered at these sites, and the correlations among
them are calculated showing good agreement with results from MC simulation.

\begin{figure}
\includegraphics{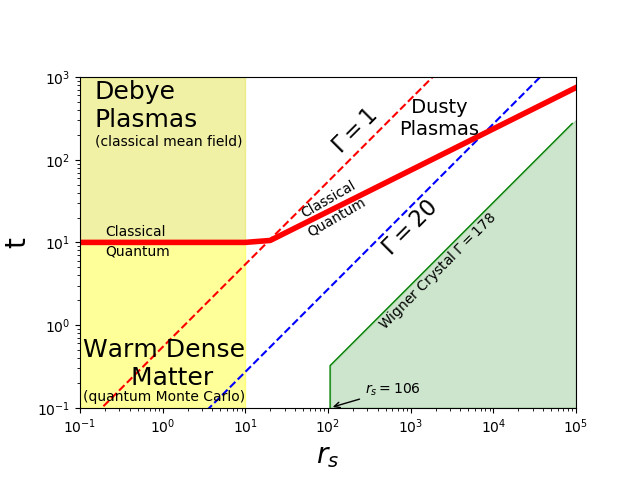}
\caption{Overview of the relevant parameter space. Here $r_s$ is the Wigner-Seitz radius in terms of the Bohr radius of the confined particles ($r_s=r_0/a_b$), and $t$ is the temperature relative to the ideal gas Fermi temperature per particles ($t=k_B T/{\varepsilon_F}$) \cite{16}. }
\label{Fig1}
\end{figure}

The effects studied here result from strong coupling conditions for which
there are relatively few theoretical methods available. In the classical
case the density functional model described below is confirmed by MD and
Monte Carlo simulations. The latter simulations are not available for the
quantum case, and the quantum density functional model has problems at
finite temperatures. However, it has been shown that the quantum system can
be mapped onto a corresponding classical system with quantum effects
embedded in effective Coulomb and trap potentials \cite{12}. Applications to
the one component plasma (jellium) show good agreement with quantum Monte
Carlo results \cite{13,14}. This approach has been applied subsequently to
the case of quantum charges in a harmonic trap \cite{15,16} as described
below. In this way a broad scope of confined Coulomb systems of interest can
be addressed. Figure \ref{Fig1} gives a simple overview of the parameter
space.

It is a pleasure to dedicate this contribution to our friend and colleague
of many years, Professor John (Jack) Sabin. Jack has been an inspiration for
all that is expected of those with an academic career, exemplifying the best
in teaching, research, and administration. His cheerful nature and good will
have brightened the lives of all who knew him well.

\section{Density functional theory}

\label{sec2}Consider $N$ particles of charge $q$ at equilibrium in a
harmonic trap at inverse temperature $\beta $. The free energy is a function
of $\beta $ and a functional of the non-uniform density $n(\mathbf{r})$, $%
F(\beta \mid n)$. The equilibrium density profile is determined from%
\begin{equation}
\frac{\delta F(\beta \mid n)}{\delta n(\mathbf{r})}=\mu -V\left( r\right) ,%
\hspace{0.25in}V\left( r\right) =\frac{1}{2}m\omega^2 r^{2}.  \label{2.1}
\end{equation}%
The potential $V\left( r\right) $ is the confining harmonic trap. The free
energy functional can be separated into that for a system without
interactions, $F_{0}(\beta \mid n)$, and a remainder $F_{ex}(\beta \mid n)$
containing all effects of Coulomb interactions among the particles%
\begin{equation}
F(\beta \mid n)=F_{0}(\beta \mid n)+F_{ex}(\beta \mid n).  \label{2.2}
\end{equation}%
A formal representation for the excess free energy in terms of pair
correlations can also be written exactly%
\begin{equation}
F_{ex}(\beta \mid n)=-\int_{0}^{1}dy\left( 1-y\right) \int d\mathbf{r}d%
\mathbf{r}^{\prime }n\left( \mathbf{r}\right) n\left( \mathbf{r}^{\prime
}\right) \beta ^{-1}c^{(2)}\left( \mathbf{r},\mathbf{r}^{\prime }\mid
y n\right) ,  \label{2.4}
\end{equation}%
where $c^{(2)}\left( \mathbf{r},\mathbf{r}^{\prime }\mid n\right) $ is the
direct pair correlation function%
\begin{equation}
\beta ^{-1}c^{(2)}\left( \mathbf{r},\mathbf{r}^{\prime }\mid n\right) \equiv
-\frac{\delta ^{2}F_{ex}(\beta \mid n)}{\delta n(\mathbf{r})\delta n(\mathbf{%
r}^{\prime })}.  \label{2.5}
\end{equation}%
In this way Eq.(\ref{2.1}) is an equation for the local density in terms of
the pair correlations of the direct correlation function \cite{9}%
\begin{equation}
\frac{\delta F_{0}(\beta \mid n)}{\delta n(\mathbf{r})}=\mu -V\left(
r\right) +\int_{0}^{1}dy\int d\mathbf{r}^{\prime }n\left( \mathbf{r}^{\prime
}\right) \beta ^{-1}c^{(2)}\left( \mathbf{r,r}^{\prime };yn\right) .
\label{2.6}
\end{equation}

The average density $\overline{n}$ is defined by%
\begin{equation}
\overline{n}=\frac{\overline{N}}{V},\hspace{0.25in}\overline{N}=\int d%
\mathbf{r}n\left( \mathbf{r}\right) .  \label{2.9}
\end{equation}%
The system is self-confined with spherical symmetry. The maximum radius $R$
is the point at which Coulomb repulsion force on a particle is balanced by
the harmonic trap confinement 
\begin{equation}
\overline{N}\frac{q^{2}}{R^{2}}=m\omega^2 R,\hspace{0.25in}V=\frac{4}{3}%
\pi R^{3}.  \label{2.10}
\end{equation}%
The mean distance between particles $r_{0}$ is introduced by 
\begin{equation}
\frac{4}{3}\overline{n}\pi r_{0}^{3}=1.  \label{2.11}
\end{equation}%
Scaling the coordinates with respect to $r_{0}$ in the above equations gives
the dimensionless form%
\begin{equation}
\frac{\delta F_{0}^{\ast }(n^{\ast })}{\delta n^{\ast }(\mathbf{r}^{\ast })}%
=\beta \mu -\frac{1}{2}\Gamma r^{\ast 2}+\int_{0}^{1}dy\int d\mathbf{r}%
^{\prime \ast }n^{\ast }\left( \mathbf{r}^{\prime \ast }\right)
c^{(2)}\left( \mathbf{r}^{\ast }\mathbf{,r}^{\prime \ast };yn^{\ast }\right)
,  \label{2.12}
\end{equation}%
where%
\begin{equation}
F_{0}^{\ast }(n^{\ast })=\beta F_{0}\left( \beta \mid n\right) ;\hspace{%
0.25in}n^{\ast }\left( \mathbf{r}^{\ast }\right) =n\left( \mathbf{r}\right)
r_{0}^{3},\hspace{0.25in}\Gamma =\beta m\omega^2 r_{0}^{2}=\beta \frac{q^2}{r_{0}%
}.  \label{2.13}
\end{equation}%
The parameter $\Gamma $ is the Coulomb coupling constant (Coulomb energy of
a pair at the average distance relative to the thermal energy $\beta ^{-1}$%
). The constant $\beta \mu $ can be eliminated in terms of $\overline{N}$.

Up to this point the results apply for both quantum and classical mechanics.
The classical case is considered more explicitly next.

\section{Classical Mechanics}

\label{sec3}Within classical statistical mechanics $F_{0}(\beta \mid n)$ can
be written exactly as a functional of the density%
\begin{equation}
F^{(0)}(\beta \mid n)=-\frac{1}{\beta }\int d\mathbf{r}n(r)\left( 1-\ln
\left( n(r)\lambda ^{3}\right) \right) .  \label{3.1}
\end{equation}%
Here $\lambda =\left( h^{2}\beta /2\pi m\right) ^{1/2}$. Then eq.(\ref{2.12}%
) becomes \cite{9} 
\begin{equation}
\ln \left( n^{\ast }(\mathbf{r}^{\ast })\right) =\ln \left( \left( \frac{%
\lambda }{r_{0}}\right) ^{3}e^{\beta \mu }\right) -\frac{1}{2}\Gamma r^{\ast
2}+\int_{0}^{1}dy\int d\mathbf{r}^{\prime \ast }n^{\ast }\left( \mathbf{r}%
^{\prime \ast }\right) c^{(2)}\left( \mathbf{r}^{\ast }\mathbf{,r}^{\prime
\ast };yn^{\ast }\right) ,  \label{3.2}
\end{equation}%
The constant first term on the right side can be eliminated in terms of $%
\overline{N}.$ Consequently the dimensional density profile and associated
free energy density profile depend only on these two parameters.

\begin{figure}
\includegraphics[width=15cm]{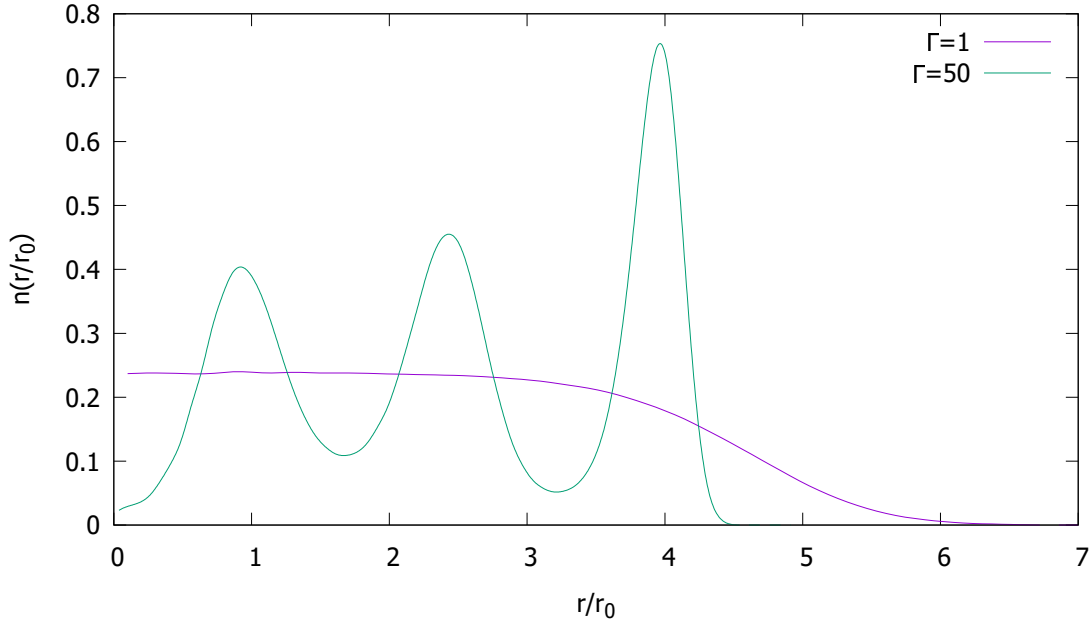}
\caption{Formation of shell structure in a harmonic trap as the coupling constant $\Gamma$ increases. Monte Carlo simulation of 100 particles in a trap.}
\label{Fig2}
\end{figure}

The solutions to (\ref{3.2}) are expected to confirm the following
qualitative behavior observed from ground state energy functions \cite{8},
and Monte Carlo and MD simulations \cite{7}. For given $%
\overline{N}$ the profiles have a strong dependence on the coupling strength 
$\Gamma $. At very small values the density profile is rotationally
symmetric and monotonically decreasing to zero from a maximum at $r^{\ast
}=0.$ At increasing values the radial dependence develops local maxima,
referred to as shells. This is illustrated in Figure \ref{Fig2}. The number
of shells increases with $\overline{N}$ and their width sharpens with
increasing $\Gamma$. The shell populations are greater for increasing
radius and grow linearly with $\overline{N}$.

Eventually, at sufficiently large $\Gamma$ rotational symmetry is broken. The
uniform distribution of particles within each shell distorts to local
domains for the associated population. Their locations are close to those of
the Thomson problem - the ground state configuration for a given number of
charges confined to a sphere. In the following the extent to which
approximations to the density functional theory (\ref{3.2}) captures these
features is described.

\subsection{Fluid phase}\label{subsec_Fluid_phase}

\begin{figure}
\includegraphics[width=15cm]{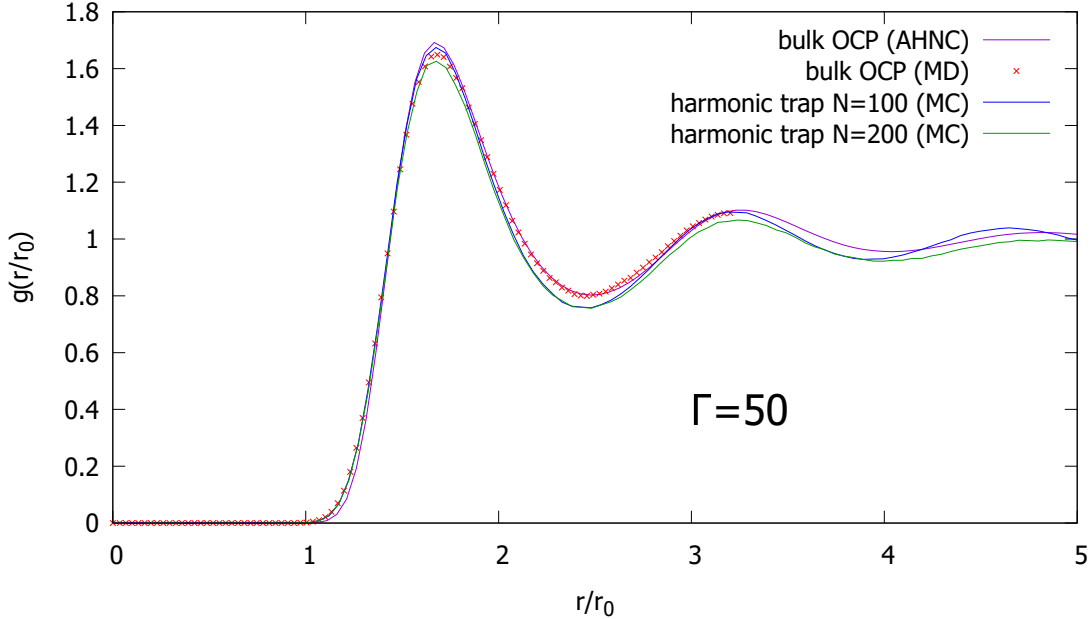}
\caption{Comparison of the pair distribution of particles in a one-component plamsa (OCP) and harmonic trap. MD calculations for the OCP were performed using Sarkas.\cite{sark1,sark2} }
\label{Fig3}
\end{figure}

\begin{figure}
\includegraphics{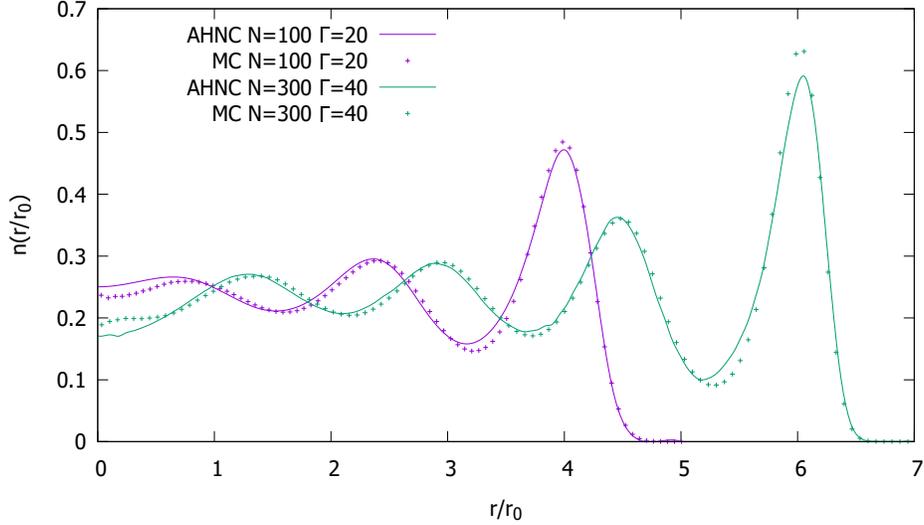}
\caption{Comparison of density profiles for AHNC and MC, for $N=100$ and $\Gamma=20$, and for $N=300$ and $\Gamma=40$.}
\label{Fig4}
\end{figure}

The approximate determination of the pair correlations described by the
direct correlation functional in (\ref{3.2}) is motivated as follows. First,
it is shown elsewhere that the functional $F_{ex}(\beta \mid n)$ is exactly
equal to the corresponding system with a uniform neutralizing background
(inhomogeneous jellium), i.e. charges in a uniform neutralizing background
with the same harmonic potential \cite{17}%
\begin{equation}
F_{ex}(\beta \mid n)=F_{jex}(\beta \mid n).  \label{3.3}
\end{equation}%
The advantage of this is that the OCP has a finite uniform limit in the
absence of the harmonic confinement, which is the uniform one component
plasma OCP.%
\begin{equation}
c^{(2)}\left( \mathbf{r,r}^{\prime };\lambda n\right) \rightarrow
c^{(2)}\left( \mathbf{r,r}^{\prime };\overline{n}\right)
=c_{OCP}^{(2)}\left( \left\vert \mathbf{r-r}^{\prime }\right\vert ;\overline{%
n}\right) .  \label{3.4}
\end{equation}%
It has been observed elsewhere \cite{11} that the distribution of pairs
within the trap \textit{without reference to their center of mass} position
are almost identical to those of the OCP, see Figure \ref{Fig3}. Therefore,
as an approximation for the fluid phase (\ref{3.4})  OCP correlations
have been used.
Evaluation of the direct correlation function for the OCP still poses a
formidable many body problem at strong coupling. However, it is a
well-studied problem and an excellent approximation, the \textit{adjusted
hypernetted chain approximation} (AHNC), is known \cite{18,19}. The solutions
to (\ref{3.2}) with these two approximations for $c^{(2)}\left( \mathbf{r,r}%
^{\prime }; n\right) $ give all of the above expected properties for
the density profile quantitatively in comparison to MD and MC results,
across the entire domain of $\Gamma $ and $\overline{N}$. An example is
illustrated in Figure \ref{Fig4}.

\subsection{Ordered phase}

\begin{figure}
\includegraphics[width=15cm]{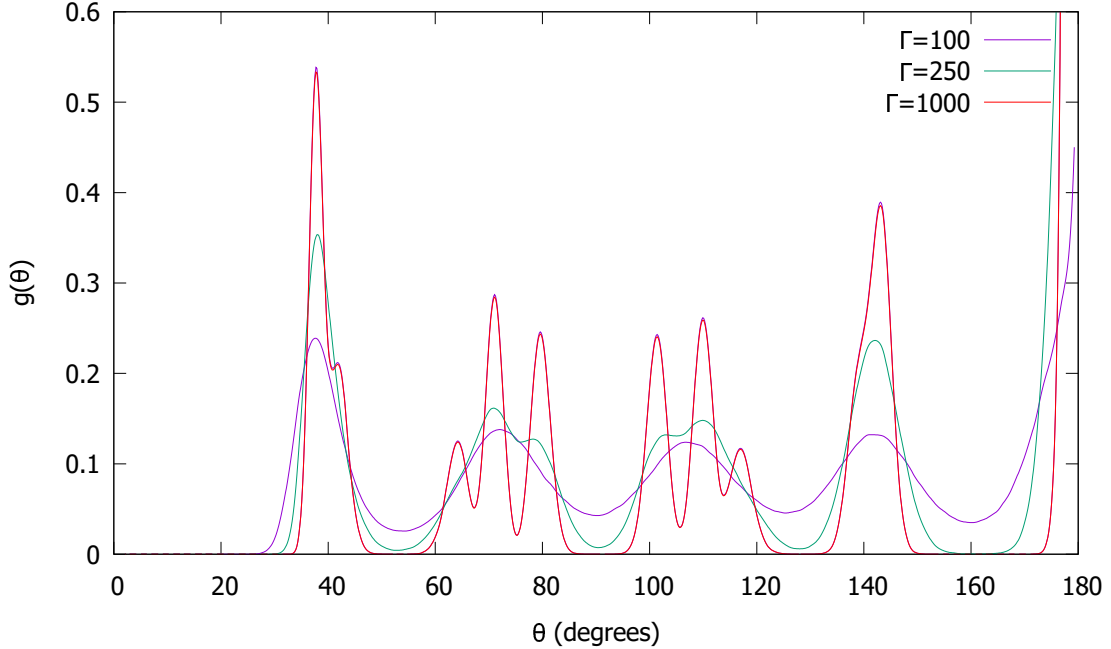}
\caption{Angular correlations within a single shell from MD simulation. The harmonic trap contained 38 particles, with 32 in the outer shell. MD simulations were performed with LAMMPS.\cite{lam1,lam2}}
\label{Fig5}
\end{figure}

In the fluid phase the particles are uniformly distributed throughout each
shell. As the coupling increases eventually the particles enter an ordered
state where rotational symmetry is broken within the shell. Figure \ref{Fig5}
shows the angular correlations within the outer shell from a molecular
dynamics simulation, for three values of the coupling constant corresponding
to fluid and ordered phases. The system consisted of $N=38$ charges, with $32
$ in the outer shell. The pair correlation function $g\left( \theta \right) $
is the probability to find a particle displaced on the shell by an angle $%
\theta $ from an arbitrary reference particle. As in a uniform fluid the
peaks represent nearest neighbor, next nearest neighbor, etc. At the lower values of $\Gamma$
 there is not much qualitative difference in the angular
correlations, but the correlation peaks are narrowing and some structure is starting to develop in the form of shoulders in the middle peaks.
For significantly larger values of $\Gamma $ however, a definite structure
appears within each peak of the correlation graph. In particular, the two broad peaks at $\Gamma =100$ that appeared at 
$70$ and $110$ degrees have condensed into triplets at $\Gamma =1000$.  In addition, the two peaks at 40 and 140 degrees are showing the
beginning formation of a doublet structure with the presence of shoulders at 
$\Gamma =1000$. This set of
doublet and triplet features persist at much higher $\Gamma$, without the appearance of any more peaks.

\begin{figure}
\includegraphics[width=10cm]{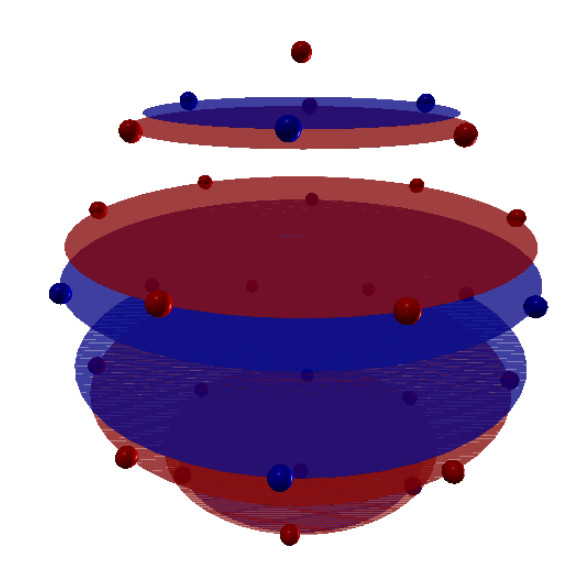}
\caption{A system of 32 charges in the Thomson problem. Horizontal circles connect those charges that are at constant angle $\theta$ from the chosen particle at the top of the figure. }
\label{Fig8}
\end{figure}

\begin{figure}
\includegraphics[width=15cm]{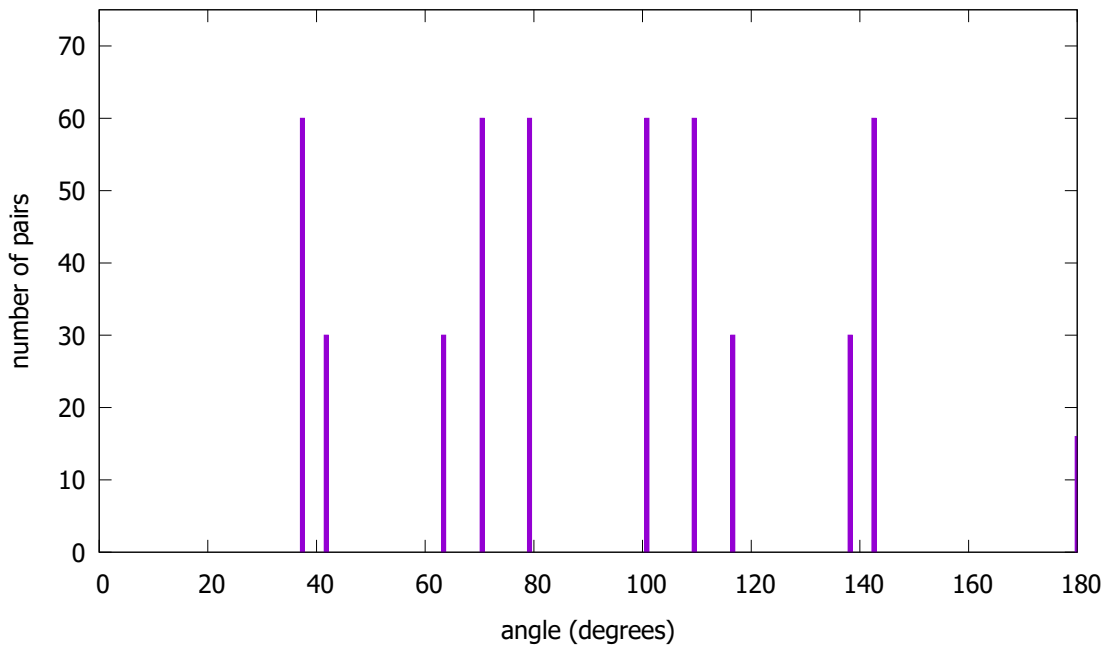}
\caption{Number of each angle present between each pair of particles in the Thomson system for $N=32$ charges.}
\label{figThompair}
\end{figure}

\begin{figure}
\includegraphics[width=15cm]{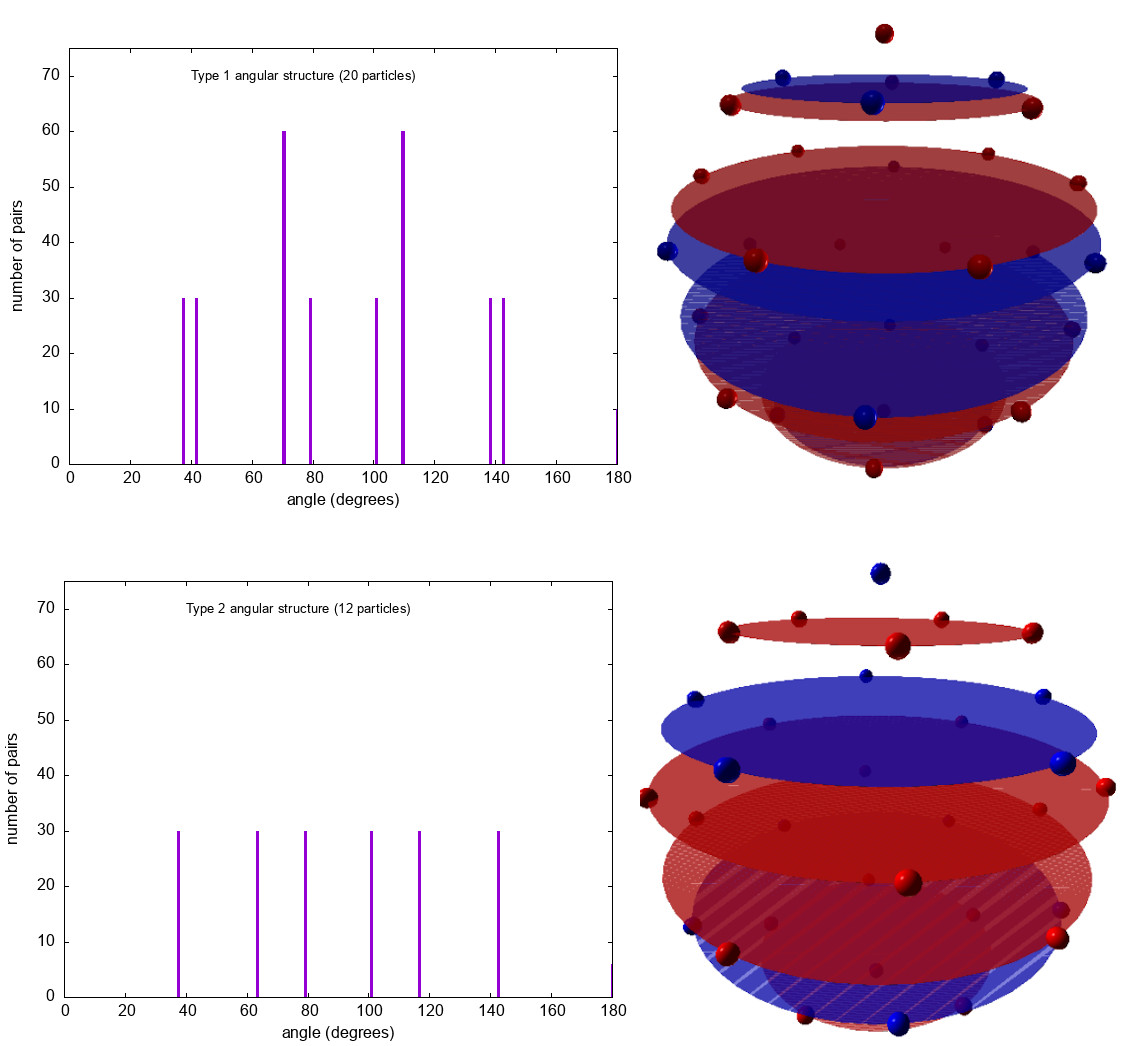}
\caption{Angular distribution for the two types of angular configuration for $N=32$ charges in the Thomson problem. The first type of angular configuration has particles at eight specific angles between $\theta=0^\circ$ and $\theta=180^\circ$. There are 20 particles in this configuration. The remaining 12 particles are in a configuration with six specific angles to the other particles. Plots show the total number of bonds that each population contributes for the entire system; their sum gives the results in Figure \ref{figThompair}.}
\label{figTwoTypes}
\end{figure}

These features can be modeled using a thin-shell model where the particle
configuration results from the Thomson problem (minimum energy configuration
for charges constrained to a sphere). The specific ordering depends on the
number of charges because of the spherical geometry. In Figure \ref{Fig8}
the Thomson configuration for a randomly chosen particle from a system containing $N=32$ charges are shown along with
constant-angle planes showing the angular displacement of the other particles.  To
compare more directly, the angular correlations of all $32$ charges in the
Thomson problem were calculated. A plot of the angular displacement of all pairs is shown in Figure \ref{figThompair}. The charges occur in two different angular correlation structures. These are shown in Figure \ref{figTwoTypes}. The combination of these two populations account for the specific angular correlation structure for the case of $N=32$ charges. 

\begin{figure}
\includegraphics{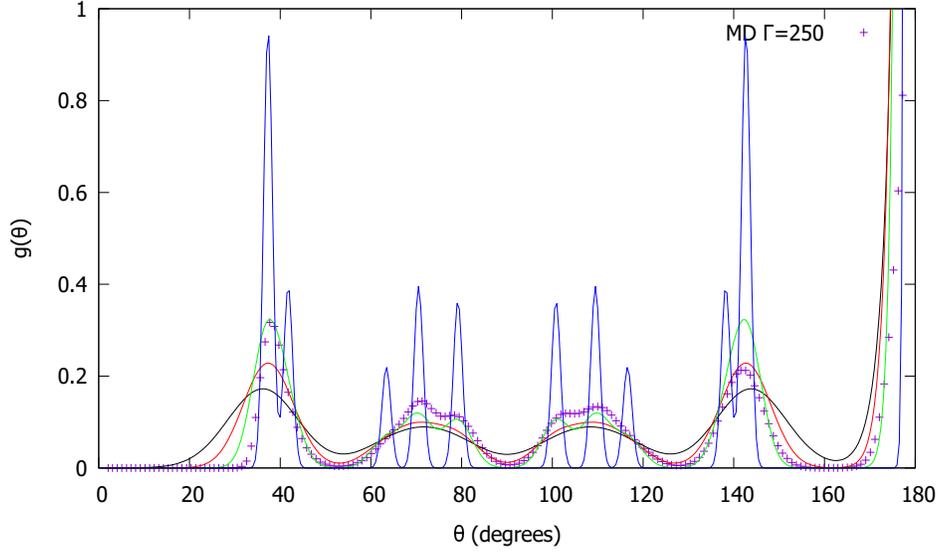}
\caption{Solid lines: evolution of the thermally-broadened angular correlations from the Thomson problem as the parameter $\alpha$ increases. Peak height increases with larger $\alpha$. Crosses: comparative results from MD at $\Gamma=250$.}
\label{Fig9}
\end{figure}

\begin{figure}
\includegraphics{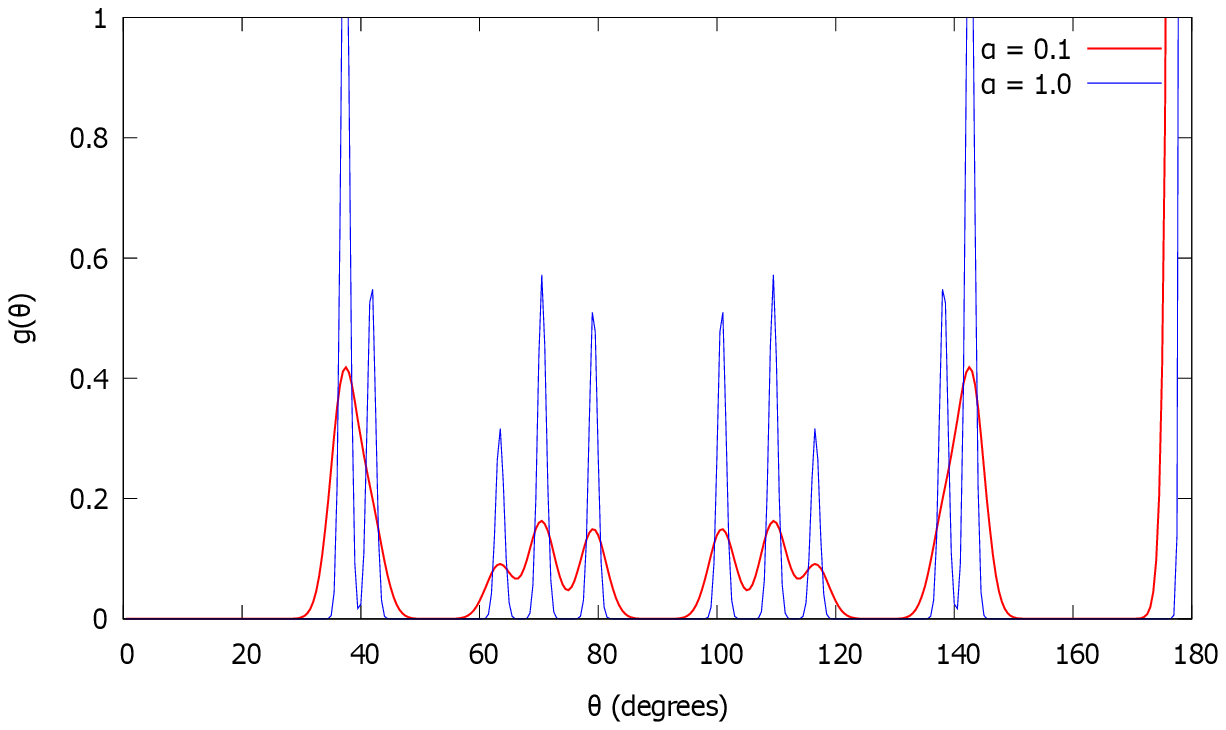}
\caption{Angular correlations from the thermally-broadened Thomson system for two values of the fitting parameter $\alpha$. Here there are $N=32$ charges in the system. The structure follows the same behavior as MD simulations for 32 particles in a shell. }
\label{Fig9b}
\end{figure}

To account for thermal effects the charges
were modeled using a Gaussian function along the sphere of the form 
\begin{equation}
f(\theta )=\sqrt{\frac{\alpha }{\pi }}\exp \left( -\alpha (\theta -\theta
_{0})^{2}\right)   \label{3.5}
\end{equation}%
where $\theta _{0}$ is the angle corresponding to the Thomson site and $%
\alpha $ is a parameter that increases with decreasing temperature. Fig \ref{Fig9} shows how this model reproduces the correct splitting, from the four
broad peaks at small $\Gamma $ which condense to the characteristic doublets
and triplets at higher $\Gamma $. Figure \ref{Fig9b} shows the effect of the  fitting parameter
 $\alpha $ showing how it models the effect of thermal motion. This
supports the idea of considering the Thomson sites on a sphere to be
analogous to a fundamental lattice for the ordered state, to the extent that
the shell can be approximated as thin.


In principle, the value of $\alpha $ in the ordered phase, for given $\Gamma
,\overline{N}$ should be obtained from minimizing the above free energy
functional using the assumed Gaussian density profile. In the fluid phase
the value of $\alpha $ would be large, representing a uniform profile. At
very large $\Gamma $, approaching the ground state, the value of $\alpha $
would approach zero. However, the assumption of the Thomson sites should
first come from solutions of (\ref{3.2}). This could be quite difficult
since ground state studies from simulations suggest there are many
metastable configurations as well.

\section{Quantum mechanics}

\label{sec4}The above classical description of confined charges at strong
coupling has exploited the methods of density functional theory, liquid
state theory, MD, and MC. At low temperatures, at or below the Fermi
temperature, quantum effects become important and many of these classical
methods do not apply directly. An accurate quantum theory at finite
temperatures, strong coupling, and confinement is still a challenging
problem. Numerical methods such as quantum Monte Carlo are applicable at
zero temperature but become less controlled as the temperature increases,
particularly for fermions. A quite different approach is to develop an exact
mapping of the quantum equilibrium structure to an effective classical
problem. This has been done recently, allowing application of the above
classical methods to quantum systems \cite{12}. Structure and correlation
calculated in this way for the OCP have shown good accuracy in comparison
with recent quantum Monte Carlo simulations \cite{13,14}.

\begin{figure}
\includegraphics{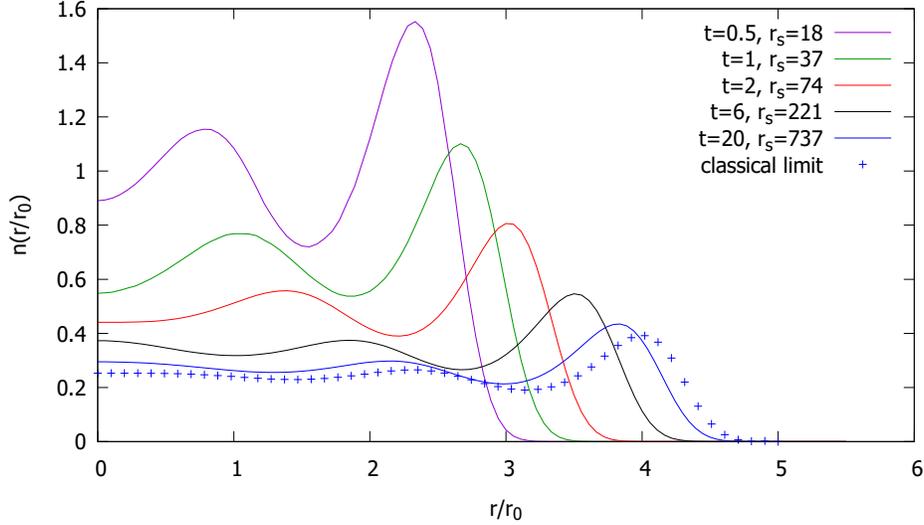}
\caption{Quantum effects for a trapped system of 100 particles. Here $\Gamma=20$ for all while the temperature parameter varies from $t=0.5$ to $t=20$. }
\label{fig10}
\end{figure}

More recently the effective classical representation of a system of quantum
charges in a harmonic trap has been explored \cite{15,16}. The primary
differences from the classical description above are modifications of the
Coulomb potential and the trap potential to accommodate quantum effects of
diffraction and exchange symmetry. A new parameter appears, in addition to $%
\Gamma $ and $\overline{N}$, the temperature relative to the Fermi
temperature ( $t=k_{B}T/e_{F}$ where $e_{F}$ is the ideal gas Fermi energy
per particle). For $t\gg 1$ the above description of shell structure and
correlations is recovered for strong coupling. At smaller $t$ the effects of
exchange degeneracy are incorporated by imposing the exact noninteracting
density profile for the ideal gas. This is non-trivial since the classical
representation of the quantum ideal gas has effective interactions. The
additional effects of exchange and diffraction are included via the direct
correlation function with modified Coulomb interactions. Figure \ref{fig10}
shows a self-similar change in the classical two shell structure being
compressed due to quantum effects on the harmonic potential.
There is much more to be done with this classical description of a quantum
system in the low $t$ domain. At much smaller $N$ connection to other
studies of quantum dots and ultracold gases should be useful. Other
properties such as spin polarization, coherent control of trap properties,
charge dependence and others acceptable to direct observation can be
addressed. A different direction for application of the results here is
obtained by the replacement of the harmonic trap with a Coulomb potential to
calculate the electron distribution about an ion. This is a solved problem
of quantum chemistry, bu its extension to a random configuration of ions is
of intense current interest for warm, dense-matter applications, e.g. the
electron density in the presence an ion configuration. Such densities are
required to compute the forces in quantum molecular dynamics simulations for
the ions in warm, dense matter at finite temperatures where traditional
density functional methods fail (e.g. the traditional Kohn-Sham
self-consistent equations for temperatures near the Fermi temperature). Here
those self-consistent equations are replaced with the classical integral
equations of AHNC. This advantage has been stressed by Dharma-wardana \cite%
{20}.

\section{Discussion}

\label{sec5}The extreme conditions of long range Coulomb charges, confined
at strong coupling and finite temperatures lead to complex structures:
radial shell structure and broken symmetry angular ordering within the
shells. At the classical conditions a density functional representation with
strong coupling correlations from the OCP is able to capture quantitatively
the transition from simple uniform filling at weak coupling to the formation
of "atomic" shell structure. At infinite coupling the ground state ordered
state is closely related to the Thomson problem (sharp shell radii, no
interactions between shells), extended here to finite temperature (e.g.,
like Debye-Waller broadening). The classical fluid phase with uniform
angular distribution is now well studied by MC, MD, and theory - the
features discussed in reference \cite{9} are given quantitatively as a
function of $\Gamma ,\overline{N}$, e.g. number of shells, occupancy,
amplitude, location. The classical ordered phase also is well studied in the
ground state, but less so at finite temperatures. The onset of localization
within shells seems not to be sharp but rather gradual, as is the formation
of shells, as a function of $\Gamma $. The Thomson sites associated with the
occupation number for a given shell provide a good reference for this
localization, as confirmed by ground state minimum energy models. The latter
models, and simulation, indicated that there are metastable configurations
with similar energy so the minimization requires care. It is possible that a
more controlled limit is obtained from the limit of finite temperature
studies as described here. 

The quantum case is well-studied at the ground state in the context of
quantum dots, nanodevices, and related systems. The case of higher
temperatures and transition to classical behavior is more limited, both from
theory and simulation. The classical map method described here is
particularly well-suited for this domain, but has not been explored very
much. Some of the advantages for states of warm, dense matter have been
outlined in reference \cite{20}. Another interesting question is the line
for Wigner crystallization (see Figure \ref{Fig1}), whose location is known
only at $T=0$ and in the high temperature classical limit.

\section{Acknowledgement}

\label{sec6}Much of the work summarized here was done in collaboration with
the Bonitz group at Christian Albrects University. The work of JW and JD was
supported by US DOE Grant DE-SC0002139.


\begin{thebibliography}{99}
\bibitem{1} J. J. Thomson, On the structure of the atom, Philos. Mag. 7, 237
(1904).

\bibitem{2} T. Pohl, T. Pattard, and J.M. Rost, Coulomb Crystallization in
Expanding Laser-Cooled Neutral Plasmas, Phys. Rev. Lett. 92, 155003 (2004).

\bibitem{3} O. Arp, D. Block, A. Piel, and A. Melzer, Dust Coulomb Balls:
Three-Dimensional Plasma Crystals, Phys. Rev. Lett. 93, 165004 (2004) ; O.
Arp, D. Block, M. Bonitz, H. Fehske, V. Golubnychiy, S. Kosse, P. Ludwig, A.
Melzer, and A. Piel, 3D Coulomb Balls: Experiment and Simulation, J. Phys.
Conf. Series 11, 234 (2005).

\bibitem{4} P. Ludwig, S. Kosse, and M. Bonitz, Structure of spherical
three-dimensional Coulomb crystals, Phys. Rev. E 71, 046403 (2005).

\bibitem{5} M. Bonitz, D. Block, O. Arp, V. Golubnychiy, H. Baumgartner, P.
Ludwig, A. Piel, and A. Filinov, Structural Properties of Screened Coulomb
Balls, Phys. Rev. Lett. 96, 075001 (2006).

\bibitem{6} V. Golubnychiy, H. Baumgartner, M. Bonitz, A. Filinov, and H.
Fehske, Screened Coulomb balls---structural properties and melting behaviour,
J. Phys. A: Math. Gen. 39, 4527 (2006).

\bibitem{7} H. Baumgartner, H. K\"{a}hlert, V. Golubnychiy, C. Henning, S. K\"{a}ding, A. Melzer, and
 M. Bonitz, Structural and dynamical properties of
Yukawa balls, Contrib. Plasma Phys. 47, 281 (2007); H. Baumgartner, D.
Asmus, V. Golubnychiy, P. Ludwig, H. K\"{a}hlert, and M. Bonitz, Ground States of Finite Spherical Yukawa Crystals, New
Journal of Physics 10, 093019 (2008).

\bibitem{8} J. Cioslowski and E. Grzebielucha, Parameter-free shell model of
spherical Coulomb crystals, Phys. Rev E 78, 026416 (2008).

\bibitem{9} J. Wrighton, J. W. Dufty, H. K\"{a}hlert, and M. Bonitz, Charge Correlations for Charges in a Harmonic Trap,
Phys. Rev E 80, 038912 (2009).

\bibitem{10} J. Wrighton, J.W. Dufty, M. Bonitz, and H. K\"{a}hlert, Shell Structure of Confined Charges at Strong Coupling, Contrib.
Plasma Phys. 50, 26 (2010).

\bibitem{11} J. Wrighton, H. K\"{a}hlert, T. Ott, P. Ludwig, H. Thomsen, J. Dufty, and M. Bonitz, Charge
Correlations in a Harmonic Trap, Contrib. Plasma Phys. 52, 45(2012).

\bibitem{12} J.W. Dufty and S. Dutta, Contrib. Plasma Phys. 52, 100 (2012);
Classical Representation of a Quantum System at Equilibrium: Theory, Phys.
Rev. E 87, 032101 (2013).

\bibitem{13} S. Dutta and J. Dufty, Classical Representation of a Quantum
System at Equilibrium: Applications, Phys. Rev. E 87, 032102 (2013).

\bibitem{14} S. Dutta and J. Dufty, Uniform electron gas at warm, dense
matter conditions, Euro. Phys. Lett. 102, 67005 (2013).

\bibitem{15} J. Wrighton, J. Dufty, and S. Dutta, Finite Temperature Quantum
Effects in Many-Body Systems by Classical Methods, in \textit{Advances in
Quantum Chemistry}, Elsevier, NY, 2015, vol 71.

\bibitem{16} J. Wrighton, J. Dufty, and S. Dutta, Finite-temperature quantum
effects on confined charges, Phys. Rev. E 94, 053208 (2016).

\bibitem{17} J. W. Dufty, Density Functional Theory for Electron Gas and for
Jellium, \textit{Langmuir} 33 (42), 11570, (2017).

\bibitem{18} J.-P. Hansen and I. MacDonald, \textit{Theory of Simple Liquids}%
, Academic Press, San Diego, CA, 1990 .

\bibitem{19} K.-C. Ng, Hypernetted chain solutions for the classical
one-component plasma up to $\Gamma$=7000, J. Chem. Phys. 61, 2680 1974.

\bibitem{sark1} L. G. Silvestri, L. J. Stanek, G. Dharuman, Y. Choi, and M. S. Murillo, Sarkas: A fast pure-python molecular dynamics suite for plasma physics, Comp Phys Comm, 272 (2022) 108245.

\bibitem{sark2} Sarkas, https://github.com/murillo-group/sarkas (2022).

\bibitem{lam1} A. P. Thompson, H. M. Aktulga, R. Berger, D. S. Bolintineanu, W. M. Brown, P. S. Crozier, P. J. in 't Veld, A. Kohlmeyer, S. G. Moore, T. D. Nguyen, R. Shan, M. J. Stevens, J. Tranchida, C. Trott, and S. J. Plimpton, LAMMPS - a flexible simulation tool for particle-based materials modeling at the atomic, meso, and continuum scales, Comp Phys Comm, 271 (2022) 10817.

\bibitem{lam2} LAMMPS, http://lammps.org (2022).

\bibitem{20} M.W. C. Dharma-wardana, A Review of Studies on Strongly-Coupled
Coulomb Systems Since the Rise of DFT and SCCS-1977, Contrib. Plasma Phys.
55, 85 (2015); M. W. C. Dharma-wardana Current Issues in Finite-T
Density-Functional Theory and Warm-Correlated Matter, Computation 4, 16
(2016).
\end{thebibliography}
\end{document}